\documentclass{PoS}

\title{RG decimation study of SU(2) gauge theory}

\ShortTitle{RG decimation study of SU(2) gauge theory}

\author{E.T. Tomboulis and \speaker{Alexander Velytsky}\\
        Department of Physics and Astronomy, UCLA, Los Angeles, CA 90095-1547, USA\\
        E-mail: \email{tombouli@physics.ucla.edu}, \email{vel@physics.ucla.edu}}

\abstract{We report on numerical studies of RG decimations in SU(2) gauge theory.
We study in particular a class of plaquette actions involving sums
of group representations. We measure a number of observables representative of different length scales
in order to investigate the transformation of the system under
different choices of spin blocking, and examine the flow of
the effective action couplings. A need for a projection to some class of effective actions
on the configurations resulting from the adopted numerical decimation procedures
is demonstrated. A numerical decimation procedure resulting in an effective
single plaquette Lagrangian tailored to reproduce different medium/large scale
physics is devised. }

\FullConference{XXIV International Symposium on Lattice Field Theory\\
		 July 23-28 2006\\
		 Tucson Arizona, US}

\begin{document}
\section{Introduction}
The Monte Carlo Renormalization Group (MCRG) technique is a useful tool for 
constructing an improved action with reduced discretization errors on the 
lattice. It has been extensively used in the search for the 
`perfect action' \cite{Hasenfratz:1997ft}, for which, under blocking transformations, trajectory 
flows approach the Wilsonian `renormalized trajectory', and lattice artifacts 
disappear. 

Under RG evolution any starting action generally develops a variety of 
additional couplings. An adequate model of the resulting effective 
action must, therefore, include a choice of several 
such couplings. In the past effective models with actions consisting of various
closed loops in the fundamental representation 
\cite{deForcrand:1999bi}, or a  
mixed fundamental-adjoint single plaquette
action \cite{Creutz:1984fj} have been studied. 
Systematic errors due to the truncation of the phase space, however, can be
significant and prevent the effective model from reaching the 
renormalized trajectory.
Also it is {\em a priori} unclear if a configuration obtained after the
numerical block spinning is representative of the equilibrium configurations 
of the adopted effective model.

After one or more block spinnings are performed, starting from a simple
(e.g. Wilson) action, one needs to measure the set of couplings retained in 
one's model of the effective action.   
This may be achieved by use of demon \cite{Creutz:1983ra} or
Schwinger-Dyson methods \cite{deForcrand:1999bi}.

Our goal is to study the long distance
confining regime of non-Abelian gauge theory. Various RG decimation schemes 
can be employed in order to connect 
perturbative short-scale with non-perturbative long scale 
physics. In Ref. \cite{Tomboulis:2005zr} an exact analytical decimating 
procedure was devised which 
imposes lower and upper bounds on such quantities as 
the partition function, and the partition function in the presence of a 
vortex, and other related quantities. The procedure can also be 
implemented numerically. Other numerical decimation schemes have 
been proposed before and are explored below. 
  
We report some progress on the construction of an effective single
plaquette action model which is designed to reproduce long/medium scale 
physics correctly. For this we 
devise special decimating procedures, which target specific medium and large 
distance observables (Wilson loops, static quark potential etc.). We pursue
both analytical and numerical procedures. Here, we report mainly on a numerical
decimation study. 

\section{Decimation procedures}
Starting with a reflection positive plaquette action, e.g.
the fundamental representation Wilson action, on lattice $\Lambda$ 
of spacing $a$, 
after $n$ decimation steps ($a\rightarrow \lambda a$) 
the partition function is 
\begin{equation}
Z_\Lambda(\beta,n)=\exp (\,\sum_{m=0}^n \phi^{(m)} |\Lambda|/\lambda^{md}\,)
\,Z_{\Lambda^{(n)}} \;,  
\end{equation}
where $\phi^{(m)}$ denotes the bulk free energy generated by each 
$\lambda^{(m-1)}a \to \lambda^{(m)}a$ step, and $Z_{\Lambda^{(n)}}$ 
is the resulting partition function at scale $\lambda^n a$. 
For a class of decimations of the `potential moving' type characterized by 
certain parameters it was shown in \cite{Tomboulis:2005zr}
that: (a) the action in $Z_{\Lambda^{(n)}}$ retains one-plaquette form, but
contains generally all representations; (b) there is a range of values of the 
parameters for which each decimation step results in an upper bound, 
$Z_\Lambda(\beta,n-1)\leq Z_\Lambda (\beta,n)$; (c) there is another range of 
parameter values for which each decimation step results in a lower bound,  
$Z_\Lambda(\beta,n-1)\geq Z_\Lambda (\beta,n)$. 
It is then possible to introduce a single parameter $\alpha$ which,
at each decimation step, interpolates between the upper and lower bound, and 
hence has a value that keeps the partition function 
constant, i.e. exact under each successive step.

The same development can be generalized to apply in presence of other 
observables, in particular vortex
free energies, i.e. twisted partition functions.
Other quantities of interest, such as Wilson loops, Polyakov loops and 
't Hooft loops, can be related to the vortex free energy through 
known inequalities. This procedure then leads to exact 
analytical results, and can also be applied as a numerical RG procedure. 
Its numerical implementation is under development. 

Here we report on numerical decimations utilizing two other 
well-known available numerical procedures. For blocking by a factor $2$ 
in all lattice directions ($a\rightarrow 2 a$), which we use throughout in 
this study, these are:  
\begin{itemize}
  \item Swendsen decimation \cite{Swendsen:1981rb}
  \begin{equation}
    Q_\mu(n)=U_\mu(n)U_\mu(n+\hat{\mu})+c\sum_{\nu\neq\mu}U_\nu(n)U_\mu(n+\hat{\nu})
    U_\mu(n+\hat{\nu}+\hat{\mu})U_{-\nu}(n+\hat{\nu}+2\hat{\mu})
  \end{equation}
  \item Double Smeared Blocking \cite{DeGrand:1994zr}
  \begin{eqnarray}
  U_\mu(n)&=&(1-6c)U_\mu(n)+c\sum_{\nu\neq\mu}U_\nu(n)U_\mu(n+\hat{\nu})
  U^\dagger_\nu(n+\hat\mu)\quad \times 2\, {\rm times}\nonumber\\
  Q_\mu(n)&=&U_\mu(n)U_\mu(n+\hat\mu).
  \end{eqnarray}
\end{itemize}
Here $c$ is the parameter which controls the relative weight of staples. 
For the Swendsen decimation $c=0.5$ and $1$ values have been used. For the 
double smeared blocking, the classical limit value $c=0.077$ has been 
used \cite{Takaishi:1995ve}.

For our  numerical decimation, we choose to start from the standard Wilson
action. After a decimated configuration is obtained, we need 
to `project' it to some effective action. Motivated by the exact 
decimation procedure, we assume that a single plaquette action 
\begin{equation}
S=\sum_{j=1/2}^{N_r}\beta_j[1-\frac1{d_j}\chi_j(U_p)],\label{eq:ef_act}
\end{equation}
truncated at some high representation $N_r$, is a general form of the 
effective action.

It is important to note that the decimated configurations may not represent 
equilibrium configurations of a
particular effective action. Therefore we follow the microcanonical
\cite{Creutz:1983ra} evolution of
the effective model starting from the decimated configurations.

\subsection{Numerical methods}
To compare the effective model to the decimated model, we need an efficient way
to simulate a gauge theory with action (\ref{eq:ef_act}). We use a procedure 
due to
Hasenbusch and Necco \cite{Hasenbusch:2004yq}. The fundamental representation 
part of the
action with specially tuned coupling is used to generate trial matrices for the
metropolis updating. This procedure typically
achieves $80\%$ acceptance rate for the metropolis algorithm at the used 
couplings. Alternatively one could use a newly developed biased metropolis 
algorithm \cite{Bazavov:2005vr}. Simple heatbath updating is 
used only in the case of purely fundamental representation action. 

For the microcanonical updating and demon measurements we implement an improved
algorithm, which retains demon energy values \cite{Hasenbusch:1994ne},
making the demon canonical.
The demons energies are restricted to $[-E_{max},E_{max}]$, thus
preventing demons from `running away' with all the energy. The couplings of the
effective action can be obtained as solutions of the equation
\begin{equation}
\langle E_d\rangle = 1/\beta
- E_{max}/tanh[\beta E_{max}].
\end{equation}
In table (\ref{tab:dem_test}) we demonstrate the ability of the canonical demon
method in measuring the couplings on $8^4$ lattice. An ensemble of 
$3000$ configurations with couplings listed in the first row of the table is
used. Demon is allowed 1 sweep for reaching equilibrium, than 10 sweeps for
measurements. The measured couplings are listed on the second row of the table
and are in good agreement with the initial values.
\begin{table}[ht]
\centering
\begin{tabular}{|c|c|c|c|c|c|}
\hline
& $\beta_{1/2}$&$\beta_1$&$\beta_{3/2}$&$\beta_2$&$\beta_{5/2}$\\\hline
in   &2.2578& -0.2201& 0.0898& -0.0333& 0.0125\\\hline
demon&2.2580(4)&-0.2206(4)&0.0903(5)&-0.0336(5)& 0.0127(4) \\
\hline
\end{tabular}
\caption{\label{tab:dem_test}Measurements of couplings by canonical demon method.}
\end{table}

\section{Decimation study}
\begin{figure}[ht]
\includegraphics[width=0.49\textwidth]{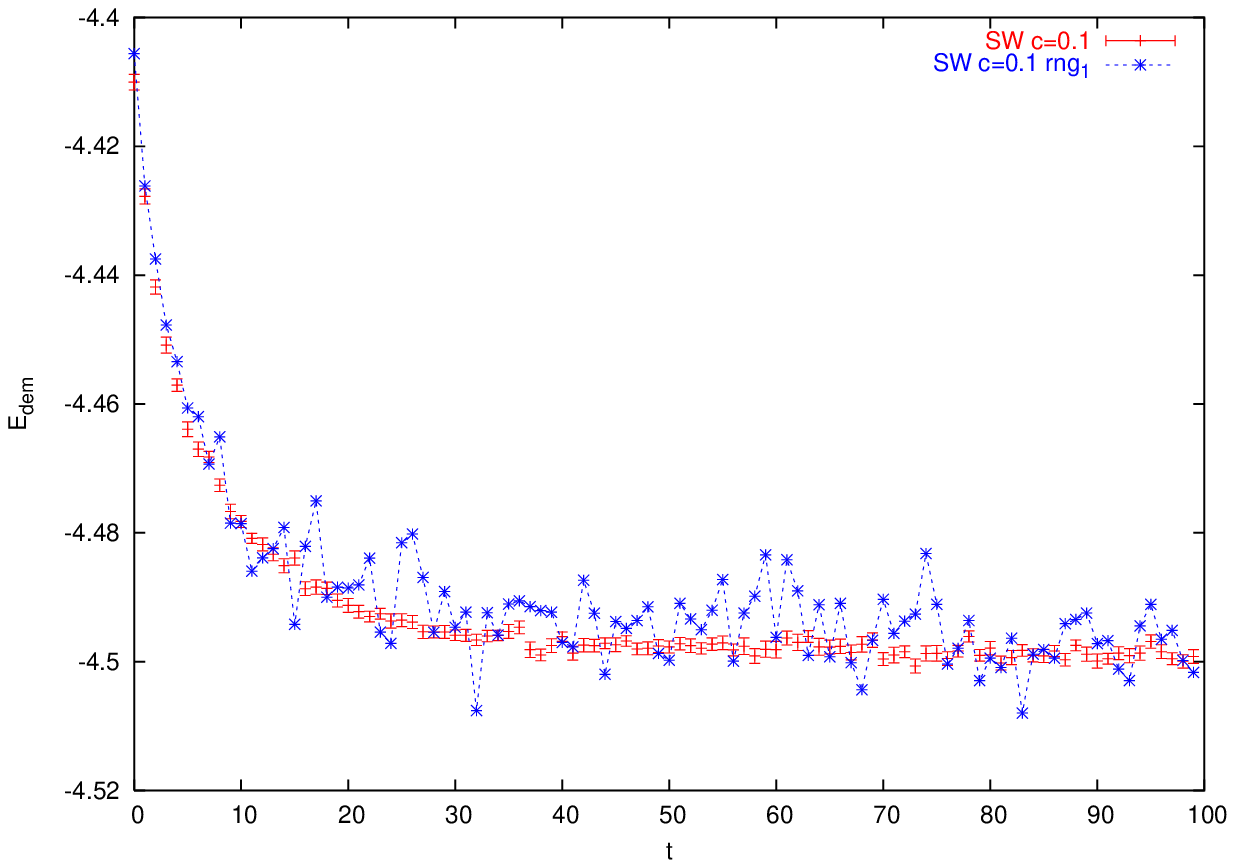}
\includegraphics[width=0.49\textwidth]{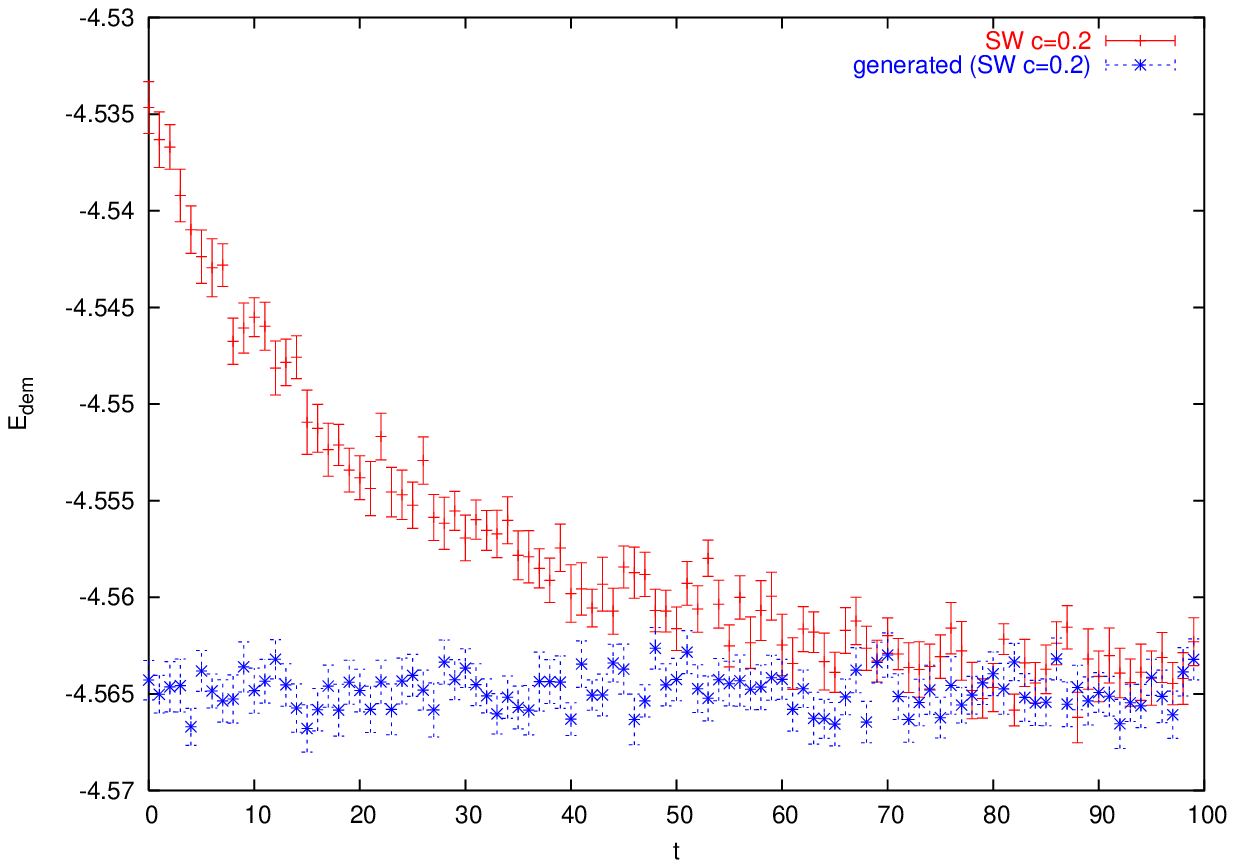}
\caption{\label{fig:d_flow}Demon fundamental energy flow for $c=0.1$ Swendsen decimated
configurations. The average and a single demon run (left).
Demon fundamental energy flow for $c=0.2$ Swendsen decimated
configurations and for configurations generated with an effective action (right).
}
\end{figure}

We fix the effective action to have 8 consecutive representations, starting
from the fundamental. A  $32^4$ lattice at $\beta=2.5$ is decimated once,
using Swendsen type decimation with various staple weights $c$. In Fig.
\ref{fig:d_flow} (left) we show the fundamental representation demon energy flow,
starting from $c=0.1$ Swendsen decimated configurations.
We note that there is a significant demon energy change during the
microcanonical evolution. The change for different replicas is always in the same
direction. There is a noticeable trend for flow stabilization at $\sim 100$
sweeps. 

Next we let the demon reach equilibrium ($>100$ sweeps) and then measure
the couplings of the effective action (\ref{eq:ef_act}). We then simulate
this model and generate thermalized configurations. We then compare
the demon evolution on these thermalized configurations with the demon
evolution on the $c=0.2$ Swendsen decimated configurations (Fig. 1 (right)). 
We see that in the former case there is no change in the demon energy, 
which indicates a very
fast demon equilibration. Whereas in the latter case there is a pronounced 
change -- this pronounced energy change is clearly due to configuration 
equilibration during the microcanonical evolution.

Therefore one concludes that starting from the (Swendsen) decimated 
configurations sufficient microcanonical evolution has to occur in order to 
`project' into the equilibrium configurations of the effective model.

\begin{figure}
\begin{minipage}[ht]{0.7\linewidth}
\includegraphics[width=0.9\textwidth]{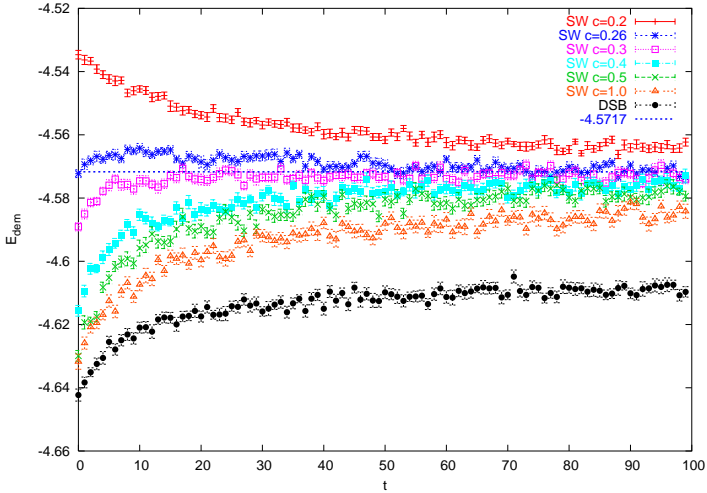}
\end{minipage}\hfill
\begin{minipage}[ht]{0.3\linewidth}
\caption{\label{fig:sw_dec_dem}Demon fundamental energy flow for various
decimated configurations (Swendsen and Double Block Smearing).}
\end{minipage}
\end{figure}

\begin{figure}
\begin{minipage}[ht]{0.7\linewidth}
\includegraphics[width=0.9\textwidth]{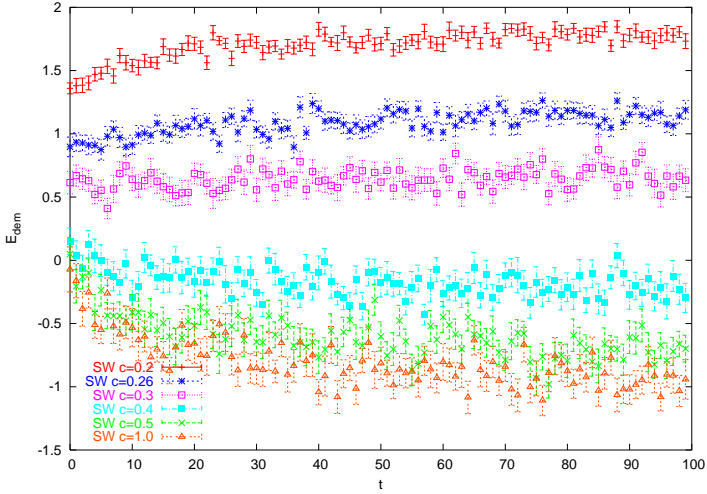}
\end{minipage}\hfill
\begin{minipage}[ht]{0.3\linewidth}
\caption{\label{fig:dem_adj}Demon adjoint energy flow for various Swendsen decimated configurations.}
\end{minipage}
\end{figure}

Next we vary the staple weight parameter $c$ and observe the demon energy
flow. In Fig. \ref{fig:sw_dec_dem} we demonstrate the fundamental demon energy
evolution for $c=0.2,\ldots,1.0$ Swendsen decimations and for the double smeared
blocking with the classical $c$ value.
We observe that there is a special $c\approx0.26$ value, when right from the
start there is little demon energy change. These particular decimation
configurations are very close to the equilibrium configuration of the model 
(\ref{eq:ef_act}).  
In Fig. \ref{fig:dem_adj} we look at the adjoint demon
energy flow. We notice that there is a small change for $c=0.26$, while
for $c=0.3$ it stays constant. This indicates that for the 
truncated actions the
decimation can produce configurations which are only approximately in 
equilibrium and the projection is generally needed.

Next we compare some medium scale physical observables measured on the decimated
configurations and on configurations obtained from the effective models. First
for the $c=0.2$ Swendsen decimation we check the difference between $N\times N$
Wilson loops measured on the decimated configuration immediately after the decimation
and after the projection of $100$ sweeps, see Tab. \ref{tab:0_100}.
\begin{table}[ht]
\centering
\small
\begin{tabular}{|c||c|c|c|c|}
\hline
$eq/m$ &$\beta_{1/2}, \beta_1, \beta_{3/2}, \ldots$ &
$\Delta P/ P^{dec}$ & 
$\Delta W_{2\times2}/ W^{dec}_{2\times2}$ &
$\Delta W_{3\times3}/ W^{dec}_{3\times3}$\\
\hline \hline
0/1&2.1391(5),-0.1628(9),0.0637(11), &&&\\
    &-0.0250(1),0.0098(15)&-0.0642(1)&-0.2832(5)&-0.7196(9)\\\hline
100/20& 2.2963(4),-0.2351(5),0.0955(7),&&&\\
    & -0.0357(9),0.0131(11),-0.0050(12)&-0.0045(1)&-0.0296(10)&-0.3912(20)\\\hline
\end{tabular}
\caption{Canonical demon measured couplings after the $c=0.2$ Swendsen decimation
and difference of various size Wilson loops estimates measured on decimated
versus effective action configurations. Measurements performed 
right after the decimation and after
100 sweeps (measurements 20 sweeps).\label{tab:0_100}}
\end{table}
We see that it is indeed the process of `projection' that makes the
difference in observables decrease.

As the next step we would like to fix the decimation parameter at the value
which minimizes the difference at the largest Wilson loop that we measure.
In Tab. \ref{tab:fix_c} we present the results for Swendsen type decimation and
Wilson loops up to $3\times3$.
\begin{table}[ht]
\centering
\begin{tabular}{|c||c|c|c|c|}
\hline
$c$ &$\beta_{1/2}, \beta_1, \beta_{3/2}, \ldots$ &
$\Delta P/ P^{dec}$ & 
$\Delta W_{2\times2}/ W^{dec}_{2\times2}$ &
$\Delta W_{3\times3}/ W^{dec}_{3\times3}$\\
\hline \hline
0.1 & 1.9912(3),-0.3085(4),&&&\\
    & 0.0990(4),-0.0362(6),&-0.0001(1)&-0.4160(6)&-0.8899(11)\\
    & 0.0139(7),-0.0045(8) &&&\\\hline
0.2 & 2.2963(4),-0.2351(5),&&&\\
    & 0.0955(7),-0.0357(9),&-0.0045(1)&-0.0296(10)&-0.3912(20)\\
    & 0.0131(11),-0.0050(12)&&&\\\hline
0.26& 2.3351(7),-0.1449(10),&&&\\
    &0.0766(12),-0.0279(13),&-0.0038(12)& 0.1502(11)&0.0926(29)\\
    &0.0084(17)&&&\\\hline
0.3 & 2.3447(8),-0.0869(12),&&&\\
    &0.0628(14),-0.0236(15),&-0.0006(1)&0.2545(12)&0.4559(41)\\
    &0.0075(20)&&&\\\hline
0.5 & 2.3618(9),0.0866(13),&&&\\
    & 0.0070(17),-0.0027(20),&0.0082(11)&0.4780(14)&1.5029(69)\\
    &-0.0013(22)&&&\\\hline
1.0 & 2.4033(9),0.1150(14),&&&\\
    &-0.0274(18), 0.0071(22),&0.0092(1)&0.4456(14)&1.4845(75)\\
    &-0.0041(29)&&&\\
\hline\hline
DS & 2.5463(11),-0.1167(17),&0.0068(1)&0.4149(14)&1.2697(70)\\
   & 0.0320(23),-0.0055(28)&&&\\
\hline
\end{tabular}
\caption{Canonical demon measured couplings after different decimations and
difference of various size Wilson loops estimates measured on decimated
and generated with effective action configurations. 
Thermalization 100 sweeps, measurements 20 sweeps.\label{tab:fix_c}}
\end{table}
We note that for different scale observables the smallest difference occurs at
different $c$ values. It is interesting that for the largest Wilson loop 
the best
results are obtained for $c=0.26$ - the value which produces configurations
closest to equilibrium. 

The classical $c$ value of double smeared blocking produces results which are
incapable of reproducing large scale physics correctly. There is obviously a 
need also in this case for a procedure similar to that described here for Swendsen type decimation.

\section*{Summary}
Multirepresentation-single plaquette  actions can  effectively 
reproduce long-scale physics. We demonstrated that a procedure naturally 
leading to a projection to some class of effective actions
is needed on the configurations resulting from any particular adopted numerical decimation scheme.
Such numerical decimation procedures applied to the effective single plaquette
Lagrangian can be tailored to reproduce different medium/large scale physics.

It is possible to extend the study by looking at the inter-quark potential. 
This would allow one to probe all length scales and check the effective
action model and decimated configurations correspondence. 
There is also a possibility to compare numerical and exact decimation
procedures. Numerically obtained coupling values can be used to connect to the
exact decimation and for consistency checks.
    
\acknowledgments
We thank Academical Technology Services at UCLA
for computer support. The simulations were 
performed on a PC cluster at UCLA. 
This work was in part supported by NSF-PHY-0309362.

\end{document}